\documentclass[12pt]{article}

\usepackage{amsfonts}
     \usepackage{amssymb}
     \usepackage{amsmath}

\def\bmx{\begin{pmatrix}}
\def\emx{\end{pmatrix}}
\def\bsq{\begin{subequations}}
\def\esq{\end{subequations}}

\def\be{\begin{eqnarray}}
\def\ee{\end{eqnarray}}
\def\bee{\begin{eqnarray*}}
\def\eee{\end{eqnarray*}}

\newtheorem{thm}{Theorem}
\newtheorem{cor}[thm]{Corollary}

\newtheorem{conj}[thm]{Conjecture}

     \def\tr{{\rm {Tr}} \, }

\def\bra{\langle}
\def\ket{\rangle}
\def\kb{ \ket \bra }

\def\ot{\otimes}

\def\b0{{\mathbf{\bold 0}}}

\title{Reliably distinguishing states in qutrit channels using
one-way LOCC}

\author{Christopher King \\
{\small      Department of Mathematics, } \\
{\small      Northeastern University,  
Boston MA 02115} 
 \\
   \and Daniel Matysiak \\
   {\small      College of Computer and Information Science,} \\
{\small Northeastern University, 
 Boston MA 02115}
}

\begin{document}

\maketitle

\begin{abstract}
We present numerical evidence showing
that any three-dimensional subspace
of ${\Bbb C}^3 \ot {\Bbb C}^n$ has an orthonormal basis which can be reliably
distinguished using one-way LOCC, where a measurement is made first on the
$3$-dimensional part and the result used to select an optimal
measurement on the $n$-dimensional part.
This conjecture has implications for the LOCC-assisted capacity of 
certain quantum
channels, where coordinated measurements are made on the system
and environment. By measuring first in the environment, the conjecture
would imply
that the environment-assisted classical capacity of
any rank three channel is at least $\log 3$. Similarly by measuring first
on the system side, the conjecture would imply that the
environment-assisting
classical capacity of any qutrit channel is $\log 3$. We also show
that one-way LOCC
is not symmetric, by providing
an example of a qutrit channel whose
environment-assisted classical capacity is less than $\log 3$.
\end{abstract}

\newpage
\section{Introduction and statement of results}
The noise in a quantum channel  arises from its interaction with
the environment. This viewpoint is expressed concisely in the
Lindblad-Stinespring representation \cite{Lind, St}:
\be\label{L-S}
\Phi(| \psi \kb \psi |) = \tr_{\cal E} \big[ U ( | \psi \kb \psi |
\ot | \epsilon \ket \bra \epsilon | ) U^{*} \big]
\ee
Here $\cal E$ is the state space
of the environment,
which is assumed to be initially prepared in a pure state $| \epsilon \ket$.
The unitary operator $U$ describes the interaction between the system
state space $\cal H$ and the environment,
and it maps product states to entangled states in ${\cal H} \ot {\cal
E}$. Taking the trace over $\cal E$ corresponds
to ignoring the environment part of the entangled state.
Since the reduced density matrix of a pure entangled state is a mixed state,
it follows that $\Phi$ generally maps pure input states to mixed output states.

\medskip

Because $U$ is unitary, orthogonal
states in the input Hilbert space $\cal H$ are mapped to orthogonal states in
${\cal H} \ot {\cal E}$.
However after the environment is traced out, the resulting output states
of the channel are no longer orthogonal, and therefore
cannot be reliably distinguished by making measurements of the channel.
If the channel is used for information transmission this
loss of distinguishability limits the
rate at which information can be reliably transmitted.

\medskip
It may be possible to more reliably distinguish output
states by using measurements on the environment in addition to
measurements on the system. This idea of using information from the
environment
to enhance channel capacity has been pursued in a
number of settings \cite{GW2, HK, Winter}. Our purpose here
is to investigate this question for some low-dimensional systems,
in order to see how
coordinated measurements on the system and the
environment
can be used to distinguish orthogonal input states.

\medskip
We will say that a set of states
$\{ \rho_i \}_{i=1}^N$ is {\em reliably distinguished} by the POVM
$\{ E_b \}_{b=1}^M$ if there is a partition
$\{1,\dots,M \} = S_1 \cup \cdots \cup S_N$ where
$S_i$ are disjoint sets,
such that $\tr \rho_i E_b = 0$ for all $b \notin S_i$.
Operationally,
this means that if the system is prepared in one
of the states $\{ \rho_i \}$, but the precise
identity of the state is unknown,
then this secret identity can be determined
by
applying the measurement $\{ E_b \}$.
In this paper we address the question:
what is the largest number of orthonormal input states
which can be reliably distinguished using one-way LOCC between the system
output and the environment?

\medskip
In order to distinguish the two directions of one-way LOCC,
we will use the notation introduced recently by Winter \cite{Winter}:
{\em environment-assisted}, meaning a measurement which is first
performed on the environment, and where the result is used to select an optimal
measurement on the system; and {\em environment-assisting},
meaning a measurement which is first
performed on the system, and where the result is used to select an optimal
measurement on the environment. More elaborate combinations
of
measurements are also defined in \cite{Winter}, but we will be concerned
only
with these two basic cases.

\medskip
\medskip
\noindent{\bf Definition 1}
{\em
For a quantum channel $\Phi$,
we denote by $N_{\{env \rightarrow sys\}} (\Phi)$ the maximal number of
orthonormal input states
which can be reliably distinguished using an environment-assisted measurement.
Similarly we denote by $N_{\{sys \rightarrow env\}}(\Phi)$ the 
maximal number of
orthonormal input states
which can be reliably distinguished using an environment-assisting
measurement.}

\medskip
Note that a given channel $\Phi$ has many possible Lindblad-Stinespring
representations of the form (\ref{L-S}). So to compute
$N_{\{env \rightarrow sys\}} (\Phi)$ and $N_{\{sys \rightarrow env\}}(\Phi)$
it is necessary to consider all such representations, and find the one
that allows the maximal number of distinguishable states.
It is known that every $d$-dimensional channel has a representation where the environment
has dimension at most $d^2$. It would be interesting to know if
$N_{\{env \rightarrow sys\}} (\Phi)$ and $N_{\{sys \rightarrow env\}}(\Phi)$
are always achieved with environments that satisfy this same bound.

\medskip 
The question of determining $N_{\{env \rightarrow sys\}} (\Phi)$ and
$N_{\{sys \rightarrow env\}}(\Phi)$ for a channel $\Phi$
with a given environment
can be re-expressed as the problem of distinguishing orthogonal
entangled bipartite
states using one-way LOCC.
Indeed, as (\ref{L-S}) shows, one way to describe the channel $\Phi$
is to specify the
subspace of entangled states $V \subset {\cal H} \ot {\cal E}$ that
are generated by the interaction, that is
\be\label{def:V}
V = \{ U (| \psi \ket \ot | \epsilon \ket) \,:\,
| \psi \ket \in {\cal H} \}
\ee
Clearly $\dim V = \dim {\cal H}$, and
there is a 1--1 correspondence between orthonormal sets in ${\cal H}$
and in $V$. So $N_{\{env \rightarrow sys\}} (\Phi)$ is the maximal number
of orthonormal
states in $V$ that can be distinguished using one-way LOCC, with measurements
first in $\cal E$ and then in $\cal H$, and vice versa for
$N_{\{sys \rightarrow env\}}(\Phi)$.
Our main result concerns orthonormal sets in
subspaces of ${\Bbb C}^3 \ot {\Bbb C}^n$, with $n \geq 3$.
Because we rely on numerical methods, we state our
results as conjectures with supporting evidence.

\medskip
\begin{conj}\label{conj1}
Let $V$ be a $3$-dimensional subspace of ${\Bbb C}^3 \ot {\Bbb C}^n$,
with $3 \leq n \leq 9$. Then there is an orthonormal basis of $V$ which can
be reliably distinguished using one-way LOCC, where measurements
are made first on ${\Bbb C}^3$, and the result used to select the optimal
measurement on ${\Bbb C}^n$.
\end{conj}

\medskip
We will present our numerical evidence for Conjecture \ref{conj1} in
Section \ref{sect.conj1}. The evidence was acquired by
sampling over
a large number of randomly selected subspaces, and
in each case
finding an orthonormal basis and a suitable local
measurement
which reliably distinguished the basis. In Appendix A we 
prove the 
following Corollary, which extends the result of
Conjecture 
\ref{conj1} to any $n \geq 3$.

\medskip
\begin{cor}\label{cor0}
If 
Conjecture \ref{conj1} holds, then the same result extends
to any 
three-dimensional subspace of ${\Bbb C}^3 \ot {\Bbb C}^n$,
for any $n \geq 3$.
\end{cor}

\medskip

Assuming that Conjecture \ref{conj1} is true, we can use it
to
find the environment-assisting capacity of any qutrit
channel.
For a qutrit channel, (\ref{def:V}) is a subspace of ${\cal
H} \ot {\cal E}$
where ${\cal H} = {\Bbb C}^3$. Hence it follows that there
is an
orthonormal basis of ${\cal H}$ whose image in ${\cal H} \ot
{\cal E}$
can be reliably distinguished by an environment-assisting
measurement.

\medskip
\begin{cor}\label{cor1}
For any qutrit channel $\Phi$,
\be
N_{\{sys \rightarrow env\}}(\Phi) = 3
\ee
\end{cor}

\medskip
We can also deduce a result whenever a channel can be
implemented
with a three-dimensional environment. The minimal
dimension of the
environment needed to implement a channel
via the Lindblad-Stinespring representation (\ref{L-S}) is
called
   the {\em  rank} of the channel. Recall that the rank  of a channel $\Phi$
which acts on states over ${\Bbb C}^d$  can be defined in several
   equivalent
ways; it is (a)
the
rank of the Choi-Jamiolkowski matrix $(I \ot \Phi) (| ME \ket \bra ME
|)$
where $| ME \ket$ is the maximally entangled state on ${\Bbb C}^d
\ot {\Bbb C}^d$,
(b) the minimal number of operators needed in a
Kraus representation for
$\Phi$, and (c) the minimal dimension of the
environment needed to implement
$\Phi$ via the Lindblad-Stinespring representation (\ref{L-S}).

\begin{cor}\label{cor2}
For any  channel $\Phi$ whose rank is three,
\be
N_{\{env \rightarrow sys\}}(\Phi) = 3
\ee
\end{cor}

\medskip
In general $N_{\{env \rightarrow sys\}}(\Phi) \neq N_{\{sys
\rightarrow env\}}(\Phi)$, and
in Section \ref{sect.rk5} we present examples of qutrit channels
for which these numbers are different. These channels all have ranks
greater than three,
   as they should by Corollary \ref{cor2}.
Interestingly we have no found no
   examples of rank four qutrit
channels for which the numbers are different, and 
  we conjecture that
   $N_{\{env \rightarrow sys\}}(\Phi) = 3$ for all rank four
qutrit channels.

\medskip

The paper is organized as follows. In Section \ref{sect.others} we
recall some related work
on the question of state discrimination using LOCC, and the recent notion of
environment-assisted capacity. In Section \ref{sect.conj1} we
describe our numerical
work in support of Conjecture \ref{conj1}, and in Section \ref{sect.rk5} we
present some examples of qutrit channels where it is not possible
to reliably distinguish three input states using environment-assisted
measurements.
Finally we discuss some conclusions in Section \ref{sect.conc}, and
the Appendix contains the proof of Corollary \ref{cor0}.

\section{Related work}\label{sect.others}
\subsection{LOCC}
There has been a lot of work on the general problem of
using LOCC (local operations and  classical communication)
to distinguish bipartite
states  \cite{GKRS, GW1, GW2}. Regarding the question of finding bases of
subspaces which can be reliably distinguished, the first result was by
Walgate et al  \cite{Wa}, who  showed that any two orthogonal bipartite states
can be reliably distinguished using LOCC.
This implies in particular that every two-dimensional subspace
of a bipartite space has a
basis that can be reliably distinguished.
It follows that $N_{\{env \rightarrow sys\}} (\Phi) = 2$  for any 
qubit channel $\Phi$
\cite{Wa, GW1}, and also that $N_{\{env \rightarrow sys\}} (\Phi) \geq
2$ for every channel \cite{HK}.

\medskip
Gregoratti and Werner \cite{GW1} have proved
the existence of bipartite subspaces which do not have any basis that can
be reliably distinguished using LOCC, though they did not provide any
explicit examples or any bounds on the dimensions of the subspaces.
Recently Watrous \cite{Wat} has constructed explicit examples of $(d^2
-1)$-dimensional
subspaces in ${\Bbb C}^d \ot {\Bbb C}^d$ which have no basis that can be
distinguished using LOCC, for $d \geq 3$. In fact Watrous' result is
even stronger,
because he proves that there is no separable POVM which can distinguish the
basis.

\medskip
More is known when constraints are put on the allowed basis vectors,
for example Nathanson \cite{Nath} has shown that
any three maximally entangled states in ${\Bbb C}^3 \ot {\Bbb C}^3$
can be distinguished using a product measurement.
Results are also known for bases composed of generalized Bell states
\cite{GKRSS}.

\subsection{Environment-assisted capacity}
In the context of channel capacities, it is natural to consider
the maximal rate at which information (classical or quantum) can be sent
though a noisy channel, using additional information gained
from
coordinated measurements of the system and environment.
Recently
Winter \cite{Winter} has analyzed the environment-assisted capacity of a
quantum channel, which is the capacity for transmission of classical
information when measurements from the environment can be
used to assist transmission. Using an asymptotic formulation of the problem
allowing entangled inputs and measurements, he has derived a lower bound
for the rate of transmission, namely half the logarithm of the dimension
of the input state space. While the results in this paper provide low-dimensional
examples where this bound is not sharp, Winter has found evidence that the lower
bound is sharp for some high-dimensional channels.

\section{Evidence for Conjecture \ref{conj1}}\label{sect.conj1}
\subsection{Partial measurements}
We implement a partial measurement
of a pure bipartite
state $| \psi \ket \in {\Bbb C}^d \ot {\Bbb C}^n$
by
projecting onto an orthonormal basis in ${\Bbb C}^d$.
If the
outcome of the
partial measurement is known, then the result is a
pure state
in ${\Bbb C}^n$. To be specific, suppose that $| v_1 \ket,
\dots, |v_d \ket$
is an orthonormal basis of ${\Bbb C}^d$, then  the
state
$| \psi \ket$ can be written in bipartite form as
\be\label{bip1}
| \psi \ket = \sum_{a=1}^d
 | v_a \ket \ot | \psi_a \ket
\ee
where $\{ | \psi_a \ket \}$
are (unnormalized) states in ${\Bbb C}^n$, and
where $\sum_{a=1}^d \bra \psi_a | \psi_a \ket = 1$. Projecting onto
the basis $\{ | v_a \ket
\}$ produces the state $| \psi_a \ket$ with probability
$\bra \psi_a | \psi_a \ket$.

\medskip
We are interested in the case where $d=3$, and
$V$
is a subspace ${\Bbb C}^3 \ot {\Bbb C}^n$.
We wish to find a basis
for $V$ and a partial measurement
on ${\Bbb C}^3$ which will
perfectly distinguish the basis states.
This will be possible if and
only if the projected states in ${\Bbb C}^n$
corresponding to each
basis state
are orthogonal for every outcome of the partial
measurement.
To be specific, let
$| \theta_1 \ket, | \theta_2 \ket ,
| \theta_3 \ket $ be an orthonormal
basis of $V$, and let $| v_1
\ket, | v_2 \ket, | v_3 \ket$ be
an orthonormal basis of ${\Bbb
C}^3$. Denote by
$| \phi(i,a) \ket$ the state in ${\Bbb C}^n$ which
results when the
state $| \theta_i \ket$ is partially projected onto
$| v_a \ket$, so that as in (\ref{bip1})
\be\label{bip2}
| \theta_i \ket = \sum_{a=1}^3  
| v_a \ket \ot | \phi(i,a) \ket
\ee
The partial measurement on ${\Bbb C}^3$ will produce
the result $a \in \{1,2,3\}$. The projected state in ${\Bbb C}^n$
will then be one of the states $| \phi(1,a) \ket, |
\phi(2,a) \ket, | \phi(3,a) \ket$, depending on which input state was used.
In order to determine the identity of the input state, it must be possible
to perfectly distinguish these three possibilities, for each measurement
result.
Therefore the basis $| \theta_1 \ket, | \theta_2 \ket , | \theta_3 \ket $
can be reliably distinguished by this partial
measurement if and only
if the three states $| \phi(1,a) \ket, |
\phi(2,a) \ket, | \phi(3,a) \ket$
are orthogonal for each
$a=1,2,3$.

\medskip
In order to test orthogonality of these states, define
\be
{\cal B} = \{ | \theta_1 \ket, | \theta_2 \ket , | \theta_3 \ket \}, \quad
{\cal M} = \{| v_1 \ket, | v_2 \ket, | v_3 \ket \}
\ee
so that $\cal B$ denotes the basis we wish to distinguish, and $\cal M$ is the
partial measurement in ${\Bbb C}^3$. We will use the following objective
function, where the various quantities are defined in (\ref{bip2}):
\be\label{def:H}
H({\cal B}, {\cal M}) = \sum_{a=1}^3 \sum_{i \neq j = 1}^3 \,
 | \bra \phi(i,a) | \phi(j,a) \ket
|^2
\ee
Clearly $H \geq 0$, and $H=0$ if and only if the
vectors $\{ | \phi(j,a) \ket \}_{j=1}^3$ are orthogonal for each
outcome $a$.

\medskip
Our goal is to find the minimal value of $H({\cal B}, {\cal M})$
for different bases in the subspace $V$ and partial measurements in ${\Bbb C}^3$.
Therefore we define
\be\label{def:H-min}
H_{\min}(V) = \min_{{\cal B}, {\cal M}} \, H({\cal B}, {\cal M})
\ee

\subsection{Random subspaces and bases}
In order to compute 
(\ref{def:H-min}) for a subspace $V$ we minimize the function $H({\cal B}, {\cal M})$
over bases of $V$ and 
bases of ${\Bbb C}^3$.
We sample
subspaces of ${\Bbb C}^3 \ot
{\Bbb C}^n$ by randomly selecting three
orthonormal vectors  $| \theta_1 \ket, | \theta_2 \ket , | \theta_3 \ket $
in ${\Bbb C}^3 \ot
{\Bbb C}^n$, and defining $V$ to be the span of these
vectors. Then every other
orthonormal basis of $V$ can be found by applying a
unitary matrix to these three vectors, that is
\be\label{def:psi-i}
| \psi_i \ket =
\sum_{j=1}^3 w_{ij} | \theta_j \ket, \qquad
i = 1,2,3
\ee
where $W =  (w_{ij})$ is a $3 \times 3$ unitary matrix.
Therefore we can search
over bases $\{| \psi_1 \ket, | \psi_2 \ket, | \psi_3 \ket \}$ of $V$ by searching over
unitary matrices
$W$.  Similarly every orthonormal basis of ${\Bbb
C}^3$ is defined by a  $3 \times 3$ unitary matrix $U$, whose columns
are the vectors in the partial measurement.

\subsection{Numerical results}
The minimization problem described above was implemented using the package TOMLAB.
Table 1 shows the results for qutrit channels organized according to
rank. Based on the results, we conjecture that all three-dimensional
subspaces of ${\Bbb C}^3 \ot {\Bbb C}^n$ have a basis which can be
perfectly distinguished using one-way LOCC, where the partial
measurement is first performed on the ${\Bbb C}^3$.

\medskip
Table 1 was obtained by randomly sampling three-dimensional subspaces of 
${\Bbb C}^3 \ot {\Bbb C}^n$, and for each subspace searching for an
orthonormal basis that could be reliably distinguished using one-way LOCC.
Column 3 shows the number of subspaces sampled for each value of $n$.
For each subspace, the objective
function (\ref{def:H}) was minimized over choices of 
orthonormal basis in $V$ and partial measurements
in ${\Bbb C}^3$. The threshold value $10^{-6}$ was used to terminate the search
for the minimum value. In every case  this threshold was reached.
Column 2 shows the average  minimum value of the objective
function when the search was terminated.

\begin{table}
    \begin{tabular}{ccc}
    \hline
    \hline
    $n$ &  Average $H$ &  \# subspaces \\
    \hline
    $3$ &  $2.816258\times10^{-8}$  & $138211$ \\
\\
    $4$ &  $3.789893\times10^{-8}$ &  $30271$  \\
\\
    $5$ &  $3.789893\times10^{-8}$  & $32278$  \\
\\
    $6$ &  $4.127063\times10^{-8}$ &  $30000$  \\
\\
    $7$ &  $4.394015\times10^{-8}$  & $30000$  \\
\\
    $8$ &  $5.130496\times10^{-8}$ &  $30216$  \\
\\
    $9$ &  $5.594670\times10^{-8}$  & $30006$ \\
    \hline
    \hline
    \end{tabular}
    \label{ndata}
    \caption{Numerical data for three-dimensional subspaces of ${\Bbb C}^3 \ot {\Bbb C}^n$
     with $n = 3, 4,\ldots, 9$, namely: the average value of the objective function  $H$ for values 
     less than the threshold value of $10^{-6}$, and the number of tested subspaces.} 
\end{table}

\subsection{One-way LOCC is not symmetric: rank five example}\label{sect.rk5}

A numerical search readily turns up examples of three-dimensional 
subspaces
of ${\Bbb C}^3 \ot {\Bbb C}^5$ which have no basis that can 
be reliably
distinguished using one-way LOCC with partial measurement 
first on the
${\Bbb C}^5$ factor. We present one of these examples in Appendix B.
Interestingly, we have found no example of a three-dimensional 
subspace of ${\Bbb C}^3 \ot {\Bbb C}^4$ which cannot be distinguished
by one-way LOCC using a partial measurement on the ${\Bbb C}^4$
factor.

\section{Conclusions}\label{sect.conc}
We have investigated the question of whether it is always possible to
reliably distinguish three orthogonal input states for a qutrit channel,
assuming that coordinated measurements between the system and
the environment are available to assist
in the state discrimination. Our numerical results indicate that the answer
is `yes', as long as measurements are first performed on the
system, and the result then used to find the optimum measurement in the
environment. The result is equivalent to the statement that every
three-dimensional subspace of ${\Bbb C}^3 \ot {\Bbb C}^n$ has a
basis which can  be distinguished by one-way LOCC, as long
as partial measurements are first performed on the ${\Bbb C}^3$ part.
This also implies that every rank three channel has a set of
three orthonormal vectors that can be reliably distinguished.
Our results
complement existing analytical results for qubit channels, and also for
the LOCC question in higher dimensional cases.

\medskip
There are several further directions to pursue in this line
of 
research. One direction
is to look for an analytical proof of our results, possibly by extending
work of Nathanson \cite{Nath} and others on LOCC state discrimination. Another
direction is to continue numerical investigations in higher 
dimensions.
We plan to work along both of these lines of 
investigation.

\bigskip
{\bf Acknowledgements} This research
was supported in part by National Science Foundation Grant
DMS-0400426. The authors are grateful to Javed Aslam
for use of computing facilities.

{~~}

\appendix

\section{Proof of Corollary \ref{cor0}}
The result of Conjecture \ref{conj1} can be formulated as a 
mathematical conjecture
concerning positive semidefinite $9 \times 9$ matrices, whose partial
trace is the identity matrix. When formulated in this way, the proof 
of
Corollary \ref{cor0} is almost immediate.

\medskip
To set up 
the notation, let $M$ be a $9 \times 9$ positive semidefinite matrix,
written in block form as
\be\label{def:M}
M = \bmx M_{11} & M_{12} & M_{13} \cr
M_{21} & M_{22} & M_{23} \cr
M_{31} & M_{32} & M_{33} \emx
\ee
and satisfying
\be\label{M1}
M_{11} + M_{22} + M_{33} = I
\ee
Letting $| \phi(i,a) \ket \in {\Bbb C}^9 = {\Bbb C}^3 \ot {\Bbb C}^3$ denote the columns of $M^{1/2}$,
it follows that
\be\label{M2}
(M_{ab})_{ij} = \bra \phi(i,a) | \phi(j,b) \ket
\ee

\medskip
Now let $\{ | v_1 \ket, | v_2 \ket, | v_3 \ket \}$ be an orthonormal
basis of ${\Bbb C}^3$ and define
\be\label{def:theta-v}
| \theta_i \ket = \sum_{a=1}^3 | v_a \ket \ot | \phi(i,a) \ket
\ee
for $i =1,2,3$. Then it follows from (\ref{M1}) that 
$| \theta_1 \ket, | \theta_2 \ket, | \theta_3 \ket$ are orthonormal.
Hence every matrix of the form (\ref{def:M}) satisfying (\ref{M1})
can be associated with a three-dimensional subspace of ${\Bbb C}^3 \ot {\Bbb C}^9$
with a chosen orthonormal basis, and with a specified basis
of ${\Bbb C}^3$.
Changing the basis vectors $| \theta_i \ket$ is equivalent to conjugating
each matrix $M_{ab}$ by the same element of $SU(3)$.
Similarly, changing measurement basis in ${\Bbb C}^3$ is equivalent to conjugating
the blocks of $M$ by a matrix in $SU(3)$. So we have the following
equivalent formulation of Conjecture \ref{conj1}.

\medskip
\begin{conj}\label{conj2}
Let $M$ be a positive semidefinite $9 \times 9$ matrix of the form
(\ref{def:M}), and satisfying (\ref{M1}).
Then Conjecture \ref{conj1} is equivalent
to the following statement: there are unitaries $W, U \in SU(3)$ such that
the diagonal blocks of the conjugated matrix
\be\label{M-diag}
(W \ot U) M (W \ot U)^{*}
\ee
commute.
\end{conj}

\medskip
We now note that any three orthonormal vectors in 
${\Bbb C}^3  \ot {\Bbb C}^n$ define a matrix of the form
(\ref{def:M}) satisfying (\ref{M1}), for any $n \geq 1$.
This follows by defining the matrix
elements of $M$ via (\ref{M2}), where $| \phi(i,a) \ket \in {\Bbb C}^n$
are defined as in (\ref{def:theta-v}). Then applying (\ref{M-diag}) 
gives the pair of unitary matrices which select the basis of $V$ and
the partial measurement in ${\Bbb C}^3$ that allow the basis
to be perfectly distinguished.

\section{Example of no one-way LOCC in ${\Bbb C}^3  \ot {\Bbb C}^5$}

Example of a subspace in ${\Bbb C}^3  \ot {\Bbb C}^5$ which
does not have a basis that can be reliably distinguished using
one-way LOCC, with partial measurement first made on the 
${\Bbb C}^5$ factor. The three states 
$\{ | \theta_1 \ket, | \theta_2 \ket, | \theta_3 \ket \}$ below generate the subspace.

\medskip
\be
\nonumber
| \theta_1 \ket  = 
\bmx
   -0.2450 - 0.0054 i \cr
   -0.1694 + 0.0815i  \cr
    0.1071 - 0.3191i \cr
    0.0655 - 0.3190i \cr
   -0.1911 - 0.1862i \cr
    0.1185 + 0.3259i \cr
   -0.2530 + 0.0480i \cr
    0.1194 - 0.1987i \cr
    0.1948 - 0.2106i \cr
    0.0595 + 0.2934i \cr
    0.1286 - 0.1427i \cr
   -0.1420 + 0.1308i \cr
   -0.2367 + 0.1399i \cr
    0.1384 - 0.0264i \cr
    0.0867 + 0.1573i  \emx, \quad
| \theta_2 \ket = \bmx
    0.1438 + 0.2108i \cr
   -0.3214 + 0.1308i \cr
    0.1229 + 0.0319i \cr
    0.1775 - 0.1070i \cr
    0.2091 - 0.1811i \cr
   -0.0937 + 0.1880i \cr
    0.1609 + 0.0272i \cr
    0.1705 + 0.0996i \cr
   -0.0630 + 0.0729i \cr
    0.3389 - 0.1242i \cr
    0.0201 - 0.2668i \cr
    0.1127 - 0.3331i \cr
    0.2338 + 0.3325i \cr
   -0.1798 - 0.0796i \cr
   -0.1097 + 0.1360i \emx 
   \ee
   \be \nonumber
| \theta_3 \ket = \bmx
    0.0390 - 0.0484i \cr
    0.0405 - 0.2603i \cr
    0.2206 + 0.2432i \cr
   -0.2843 - 0.0751i \cr
   -0.2416 - 0.1380i \cr
    0.0510 + 0.3270i \cr
    0.1691 + 0.0829i \cr
   -0.3761 - 0.1033i \cr
    0.0988 + 0.1388i \cr
    0.3138 + 0.2228i \cr
    0.0553 + 0.2272i \cr
    0.0468 - 0.0164i \cr
    0.1966 - 0.1044i \cr
   -0.0147 + 0.1239i \cr
   -0.2313 + 0.0715i  \emx
\ee

\end{document}